\documentclass{ws-ijmpa}
\usepackage[super,compress]{cite}
\usepackage{graphicx}
\begin{document}
\markboth{N. Khusnutdinov}{Self-force and the Huygens principle}

%
\catchline{}{}{}{}{}
%
\title{Self-force and the Huygens principle}
\author{N. Khusnutdinov}
\address{Centro de Matem\'atica, Computa\c{c}\~ao e Cogni\c{c}\~ao, Universidade Federal do ABC, 09210-170 Santo Andr\'e, SP, Brazil, and \\Institute of Physics, Kazan Federal University, Kremlevskaya 18, Kazan, 420008, Russia \\ nail.khusnutdinov@gmail.com}

\maketitle

\begin{history}
	\received{Day Month Year}
	\revised{Day Month Year}
\end{history}

\begin{abstract}
	We consider a relation between the Huygens Principle (HP) in gravity and the self-interaction force. We show that the self-force for an electric particle in the plane gravitational wave space-time has no tail term even the vector Green function does not obey the HP. The reason for this observation is that even vector potential does not obey the HP, the electromagnetic field does obey. 
\end{abstract}  

\ccode{PACS numbers: 04.30.-w, 03.50.De}	
	
\section{Introduction} 

The Huygens Principle (HP), originated in 1690 was a very powerful principle in physics. Hadamard  has formulated this principle in the following form \cite{Hadamard:1923:locpilpde} (see also Ref. \citen{Baker:1939:tmtohp}) 
\begin{description}
	\item[(A) (major premise)] ``The action of phenomena produced at the
instant $t = 0$ on the state of matter at the later time $t = t_0$ takes place by the mediation of every intermediate instant $t=t'$, i.e. (assuming $0 < t' < t_0$), in order to find out what takes place for $t = t_0$, we can deduce from the state at $t = 0$ the state at $t=t'$ and, from the latter, the required state at $t = t_0$."
	\item[(B) (minor premise)] ``If, at the instant $t = 0$ -- or more exactly throughout a short interval $- \epsilon \leq t \leq 0$ -- we produce a luminous disturbance localized in the immediate neighborhood of $0$, the effect of it will be, for $t = t'$, localized in the immediate neighborhood of the
	surface of the sphere with center $0$ and radius $\omega t'$: that is, will be localized in a very thin spherical shell with center $0$ including the
	aforesaid sphere."
	\item[(C) (conclusion)] ``In order to calculate the effect of our initial luminous phenomenon produced at $0$ at $t = 0$, we may replace it by a 	proper system of disturbances taking place at $t = t'$ and distributed over the surface of the sphere with center $0$ and radius $\omega t'$."
\end{description}

The HP is valid if the fundamental solution of the wave equation is concentrated on the null surface, the sphere of radius $|\boldsymbol{r}| = t$. In flat space-time the Principle is valid for space dimension $d=3$ and violated for space dimension $d=2$. Indeed, for $d=3$ the fundamental solution of scalar wave equation\footnote{Here, $\overline{D}_n = (D^{ret}_n + D^{adv}_n)/2$} $\overline{D}_3 = \delta (\sigma(x,x'))/8\pi$ and for $D=2$: $\overline{D}_2 = \theta (\sigma(x,x'))/4\pi\sqrt{2\sigma (x,x')}$, where $\sigma(x,x') = (t^2 - \boldsymbol{r}^2)/2$ -- half of the square geodesic interval between points $x$ and $x'$ and $\theta (x)$ is a step function (see, for example Ref. \citen{Vladimirov:1971:eomp}).  For $d=3$ the wave field is concentrated on the null surface $\sigma = 0$ while for $d=2$ -- not. 

In curved space-time the situation is different. For $d=3$ the vector Green function has the following form  \cite{DeWitt:1965:DTGF} 
\begin{equation}\label{eq:a}
\overline{D}_{\mu\nu'}   = \frac{\sqrt{\Delta}}{8\pi} \textsl{g}_{\mu\nu'} \delta (\sigma) +  \frac{\sqrt{\Delta}}{8\pi} \theta (-\sigma) a_{\mu\nu'},
\end{equation}
where $\Delta$ is the van Fleck--Morette determinant, $\textsl{g}_{\mu\nu'}$ is bivector of parallel transport along a geodesic. We observe that the Green function has part concentrated inside the null surface alongside with part which survives on the null surface.  Therefore, in general, HP is violated in curved space-time. It was a great activity about proving the validity of Principle in different Petrov's classes of space-times (see, for example Refs. \citen{Sonego:1992:hpacppfwicst,McLenaghan:1996:noptistowwneomeshp,Anderson:1999:hpftnsasweoptist,Czapor:1999:csohpftsweoptist}). 

\section{The Self-force}

The Green function plays a great role in the description of another phenomenon -- self-interaction. Considering that the electromagnetic field is an objective reality with own energy and momentum we have to take into account its interaction with electromagnetic particles. The phenomenon of self-force in Minkowski space-time was considered in details in monographs \cite{Sokolov:1986:RRE,Landau:1975:CTF} and reviews.  \cite{Klepikov:1985:rdfarfcp,Krivitskii:1991:arrfiqe} Equation of motion with self-force contribution reads  
\begin{equation}
m\frac{Du^\mu}{d s}= eF^{\mu\nu}_{ext} u_\nu + \frac{2e^2}{3}\frac{D^2u^\nu}{ds^2}P^\mu_\nu, 
\end{equation}
where $P_{\mu\nu} = \textsl{g}_{\mu\nu} + u_\mu u_\nu$ is the projector on the velocity of the particle. 

In general relativity, the situation with self-interaction force becomes more complicated.  \cite{Poisson:2004:tmoppics,Khusnutdinov:2005:Psegf,Poisson:2011:tmoppics} The equation of motion with self-force contribution has the following form \cite{DeWitt:1960:rdiagf,Hobbs:1968:avford} 
\begin{equation}
m\frac{Du^\mu}{d s}= \frac{2e^2}{3}\frac{D^2u^\nu}{ds^2}P^\mu_\nu +\frac{e^2}{3}R^\nu_\beta u^\beta P^\mu_\nu + e^2 u^\alpha \int_{-\infty}^s f^\mu_{\,\cdot\alpha\beta'}u^{\beta'} d s',\label{eq:SFGeneral}
\end{equation}
where $f_{\mu\alpha\beta'} =  v_{\mu\beta',\alpha} - v_{\alpha \beta',\mu}$ and $v_{\mu\nu'} = \sqrt{\Delta} a_{\mu\nu'}/8\pi$. The self-force has a non-local contribution which is defined by that part of the Green function which violates the HP. If $v_{\mu\nu'} \not = 0$, the HP is violated and non-local contribution in self-fore appears. But there is another possibility when $v_{\mu\nu'} \not = 0$ but   $f_{\mu\alpha\beta'}=0$. This situation is realized in the plane gravitational wave (GW) space-time with metric  
\begin{equation}\label{eq:metric}
ds^2 = - 2du dv + \gamma_{ij}(u) dx^i dx^j; \ u,v = (t \mp x)/\sqrt{2}.
\end{equation} 
Straightforward calculations give the following expression for $v_{\mu\nu'} $:
\begin{equation}\label{eq:v}
v_{\mu\nu'}(u,u') =   \delta_\mu^u \delta_{\nu'}^{u'} \frac{\sqrt{\Delta}}{8\pi \delta u}  \left(\delta ( \ln L^2)\dot{} + p_{ij} \delta \lambda^i_{(a)} \delta \lambda^{j(a)}\right),
\end{equation}
where $\Delta = \delta u^2 \det p_{ij}/L^2(u) L^2(u')$, $L^2 = \det \gamma_{ij}$, $\delta f = f(u) - f(u')$, and 
\begin{equation}\label{eq:p}
	p^{ij} = \int_{u'}^{u} \gamma^{ij}du,\ p_{ij}p^{jk} = \delta^k_i. 
\end{equation}
Here we used the following frame 
\begin{equation}
	\lambda^\mu_{(a)}  = 
	\left(\begin{array}{cccc}
		1&0&0&0\\
		0&1&0&0\\
		0&0&\dfrac{L^2 \cos \psi  - \gamma_{23} \sin\psi}{\sqrt{\gamma_{22}}L^2}& \dfrac{\sqrt{\gamma_{22}} \sin\psi}{L^2}\\
		0&0&-\dfrac{\gamma_{23}\cos\psi + L^2\sin \psi }{\sqrt{\gamma_{22}}L^2}& \dfrac{\sqrt{\gamma_{22}} \cos\psi}{L^2}
	\end{array}\right),
\end{equation}
with specific expression for angle 
\begin{equation}
	\psi = \int^u_{u'} \frac{\gamma_{23} \dot{\gamma}_{22} - \dot{\gamma}_{23} \gamma_{22}}{2\gamma_{22} L^2}du.
\end{equation}

The bivector $v_{\alpha\beta'}$ depends on $u$ and $u'$, only and it has the following structure $v_{\alpha\beta'} = \delta_\alpha^u \delta_{\beta'}^{u'} q(u,u')$ (see Eq. (\ref{eq:v})). For this reason $f_{\mu\alpha\beta'} = v_{\mu\beta',\alpha} - v_{\alpha \beta',\mu} = 0$. Therefore, we conclude that the tail part of self-force is zero and the self-force has local form even the vector Green function does not obey HP. For a scalar particle, the tail  part of self-force is zero because the Green function has local form. \cite{Gibbons:1975:qfpipws}

The point is that the electromagnetic field obeys HP,  even the vector potential does not. Indeed, the non-local term (\ref{eq:v}) in the vector Green function  gives a contribution to the following components of the Maxwell tensor $(a=v,y,z)$
\begin{equation}\label{eq:Huygens}
F_{ua} (x) = -4\pi e \left.\frac{\sigma_{,a}(x,x(s))u_v(s)}{\frac{d\sigma(x,x(s))}{ds}}  v_{uu'}(u,u(s))\right|_{s\to s^*},
\end{equation}
where retarded time $s^*$ is the solution of the equation $\sigma (x,x(s^*)) = 0$ and $u(s^*) \leq u$. Therefore, we may affirm, that the HP is valid for an electromagnetic field while it is not obeyed for the vector potential of the electromagnetic field. Firstly, it was noted by K\"unzle in Ref. \citen{Kunzle:1968:mfshp} (see, also discussion in Ref. \citen{Sonego:1992:hpacppfwicst}).

\section{Conclusion}

We considered a connection between HP and non-local contribution to the self force. Obviously, if the Green function does not obey HP (contains tail part $a_{\mu\nu'}$ \eqref{eq:a}) then the non-local contribution in self-force exists (last term in Eq. \eqref{eq:SFGeneral}). In the same time, there is another possibility for absence tail part in self-force, namely when $f_{\mu\alpha\beta'} =  v_{\mu\beta',\alpha} - v_{\alpha \beta',\mu} = 0$, where $v_{\mu\nu'} = \sqrt{\Delta} a_{\mu\nu'}/8\pi$. We have shown that this situation is realized in the case of plane GW background with metric \eqref{eq:metric}. In this case, the vector Green function does not obey to HP, but the electromagnetic field does.  This point was noted firstly by K\"unzle. \cite{Kunzle:1968:mfshp} 

\section*{Acknowledgments}
I'd like to thank G. Gibbons for helpful discussions on HP in gravity. The author was supported in part by the Russian Foundation for Basic Research Grant No. 19-02-00496-a and by the grants 2019/10719-9, 2019/06033-4 of S\~ao Paulo Research Foundation (FAPESP).


\begin{thebibliography}{10}
	
	\bibitem{Hadamard:1923:locpilpde}
	J.~Hadamard, {\em {Lectures on Cauchy's Problem in Linear Partial Differential
			Equations}} (Yale University Press, New Haven, 1923).
	
	\bibitem{Baker:1939:tmtohp}
	B.~B. Baker and E.~T. Copson, {\em {The Mathematical Theory of Huygens
			Principle}} (Oxford University Press, Oxford, 1939).
	
	\bibitem{Vladimirov:1971:eomp}
	V.~S. Vladimirov, {\em {Equations of Mathematical Physics}} (Marcel Dekker
	Inc., New York, 1971).
	
	\bibitem{DeWitt:1965:DTGF}
	B.~S. DeWitt, {\em {Dynamical Theory of Groups and Fields}} (Gordon and Breach,
	New York, 1965).
	
	\bibitem{Sonego:1992:hpacppfwicst}
	S.~Sonego and V.~Faraoni, {\em J. Math. Phys.} {\bf 33}, 625  (1992).
	
	\bibitem{McLenaghan:1996:noptistowwneomeshp}
	R.~G. McLenaghan and F.~D. Sasse, {\em Ann. Henri Poincar\'e} {\bf 65}, 253
	(1996).
	
	\bibitem{Anderson:1999:hpftnsasweoptist}
	W.~G. Anderson, R.~G. McLenaghan and F.~D. Sasse, {\em Ann. Henri Poincar\'e}
	{\bf 70}, 259  (1999).
	
	\bibitem{Czapor:1999:csohpftsweoptist}
	S.~R. Czapor, R.~G. McLenaghan and F.~D. Sasse, {\em Ann. Henri Poincar\'e}
	{\bf 71}, 595  (1999).
	
	\bibitem{Sokolov:1986:RRE}
	A.~A. Sokolov and I.~M. Ternov, {\em {Radiation from Relativistic Electrons}}
	(AIP, New York, 1986).
	
	\bibitem{Landau:1975:CTF}
	L.~D. Landau and E.~M. Lifshitz, {\em {The Classical Theory of Fields}}
	(Pergamon Press, Oxford, 1975).
	
	\bibitem{Klepikov:1985:rdfarfcp}
	N.~P. Klepikov, {\em Sov. Phys. Usp.} {\bf 28},   506  (1985).
	
	\bibitem{Krivitskii:1991:arrfiqe}
	V.~Krivitskii and V.~Tsytovich, {\em Sov. Phys. Usp.} {\bf 34}, 250  (1991).
	
	\bibitem{Poisson:2004:tmoppics}
	E.~Poisson, {\em Living Rev. Relativ.} {\bf 7}, 1  (2004).
	
	\bibitem{Khusnutdinov:2005:Psegf}
	N.~R. Khusnutdinov, {\em Phys. Usp.} {\bf 48}, 577  (2005).
	
	\bibitem{Poisson:2011:tmoppics}
	E.~Poisson, A.~Pound and I.~Vega, {\em Living Rev. Relativ.} {\bf 14}, 1
	(2011).
	
	\bibitem{DeWitt:1960:rdiagf}
	B.~DeWitt and R.~Brehme, {\em Ann. Phys.} {\bf 855}, 220  (1960).
	
	\bibitem{Hobbs:1968:avford}
	J.~M. Hobbs, {\em Ann. Phys.} {\bf 47}, 141  (1968).
	
	\bibitem{Gibbons:1975:qfpipws}
	G.~W. Gibbons, {\em Commun. Math. Phys.} {\bf 45}, 191  (1975).
	
	\bibitem{Kunzle:1968:mfshp}
	H.~P. K\"unzle, {\em Math. Proc. Cambridge Philos. Soc.} {\bf 64},   779
	(1968).
	
\end{thebibliography}
\end{document}